\documentclass[useAMS,usenatbib]{mn2e}

\title{Near IR Spectroscopic monitoring of WR 140 during the 2001 periastron passage}
\author[Watson P. Varricatt, P. M. Williams \& N. M. Ashok]
       {Watson P. Varricatt$^{1}$\thanks{E-mail: w.varricatt@jach.hawaii.edu(WPV);
        pmw@roe.ac.uk(PMW); ashok@prl.ernet.in(NMA)},
        P. M. Williams$^{2}\footnotemark[1]$ \& N. M. Ashok$^{3}$\footnotemark[1]  \\
   $^{1}$Joint Astronomy Centre, 660 N. Aohoku Place, Hilo,  Hawaii 96720, USA\\
   $^{2}$Institute for Astronomy, University of Edinburgh, Royal Observatory, Edinburgh EH9 3HJ, UK\\
   $^{3}$Physical Research Laboratory, Navrangpura, Ahmedabad, India 380009}

\date{This is a preprint of the article accepted for publication in MNRAS {\copyright}[2004], The Royal Astronomical Society - 
      Received 2003 December;}

\pagerange{\pageref{firstpage}--\pageref{lastpage}}
\pubyear{2004}

\begin{document}

\maketitle

\label{firstpage}

\begin{abstract}

We present new spectra of WR 140 (HD 193793) in the $JHK$ bands with some
covering the 1.083-$\mu$m He\,{\sc i} emission line at higher resolution,
observed between 2000 October and 2003 May to cover its 2001 periastron
passage and maximum colliding-wind activity.  The WC7 + O4-5 spectroscopic
binary WR 140 is the prototype of colliding-wind, episodic dust-making
Wolf-Rayet systems which also show strong variations in radio and X-ray emission.
The $JHK$ spectra showed changes in continuum and in the equivalent widths of
the WC emission lines, consistent with formation of dust starting between 
2001 January 3 and March 26 (orbital phases 0.989 and 0.017) and its 
subsequent fading and cooling.  The 1.083-$\mu$m He\,{\sc i} line has a 
P-Cygni profile which showed variations in both absorption and emission 
components as WR 140 went through periastron passage. The variation of the
absorption component of the profile yielded tight constraints on the geometry 
of the wind-collision region, giving $\theta = 50^{\circ}\pm8^{\circ}$ 
for the opening semi-angle of the interaction region `cone', indicating 
a wind-momentum ratio 
$(\dot{M} v_{\infty})_{\rm O} / (\dot{M} v_{\infty})_{\rm WR} = 0.10$,  
about three times larger than previously believed.  
As the system approached periastron, the normally flat-topped emission 
component of 1.083-$\mu$m line profile showed the appearance of a 
significant sub-peak. The movement of the sub-peak across the profile was 
seen to be consistent with its formation in wind material flowing along 
the contact discontinuity between the WC7 and O4-5 stellar winds and the 
changing orientation of the colliding wind region as the stars moved in 
their orbits. The flux carried in the sub-peak was significant, exceeding 
the X-ray fluxes measured at previous periastron passages. This additional 
source of radiative cooling of the shock-heated gas probably causes it to depart
from being adiabatic around periastron passage, thereby accounting for the 
departure of the X-ray flux from its previously expected $1/d$-dependency.
\end{abstract}


\section{Introduction}

Stars with  M$_{\ast} \geq$ 30M$_{\odot}$ go through a Wolf-Rayet (WR)
phase in their evolution, during which they undergo large scale mass loss
($\dot{M} \sim$ 10$^{-5}$ $M_{\odot}$ year$^{-1}$).  Consequently they 
display broad emission line spectra and large infrared excesses 
originating in their accelerated stellar winds (wind terminal velocity 
v$_{\infty}$=750--5000 km s$^{-1}$). They are post-main sequence stars 
which have lost most of their hydrogen envelopes through stellar winds
in the pre-WR stage. Based on the emission lines seen, they are classified
as WN stars (lines of He, N and H), WC and WO stars (lines of He, C and O).
Late WC stars (WC7--9) show evidence of circumstellar dust, mainly of 
amorphous carbon (Williams 1995).  Some of them are persistent dust makers
which show continuous dust formation and others are episodic dust makers, 
where we see periodic rises in their IR flux due to episodic formation of
dust. It is believed, now, that episodic dust making WC stars are massive
binary systems in which the stellar winds of two massive stars interact.
WR 140 (HD 193793) is the prototype of colliding-wind, episodic dust 
forming WC stars (Williams 1995). 

From the photometry of the infrared maxima in 1977 and 1985 and earlier 
data, Williams et al. (1987) found that the occurence of the IR maxima 
and the dust formation were periodic and derived a period of 7.9 years. 
Studying the existing radial velocity data, from which no orbit had 
previously been derived, they showed that WR 140 was indeed a spectroscopic 
binary with high eccentricity. This was extended to a multi-wavelength 
study, including 1.65--12.6 $\mu$m light curves, radio and X-ray data 
by Williams et al. (1990, hereafter W90), who derived a photometric period 
$P=2900 \pm 10$ days (7.94 y) and a spectroscopic orbit from published radial 
velocities giving elements $e=0.84\pm0.04$, $\omega$=32$^{o}\pm$8$^{o}$ and 
T$_{0}$=JD 2446160$\pm$29 (1985.26$\pm$0.08)). The system formed dust for 
about four months, 4\% of the period, during periastron passage. This is 
perhaps one of the least well understood of the phenomena shown by WR 140. 

The most recent periastron passage and dust-formation episode occurred 
in early 2001, for which a multi-wavelength observing campaign was 
planned and undertaken. 

Marchenko et al. (2003) carried out optical photometric and spectroscopic 
observations and derived a spectroscopic period $P=2899\pm1.3$ days, in 
excellent agreement with the photometric period and new orbital 
elements, $e$=0.881$\pm$0.005, $\omega$=46.7$^{o}\pm$1.6$^{o}$ and 
T$_{0}$=JD 2446147.4$\pm$3.7, which we use in this paper.  They found
that at phases very close to periastron,  the profiles of the 
5696\AA  $\>\>$C\,{\sc iii} and 5876\AA $\>\>$He\,{\sc i} lines developed extra 
emission components at their `blue' ends which rapidly moved to the red. 
 
Panov \& Dinko (2002) observed a  minimum in UBV photometric 
bands in 2001 June (phase = 0.038-0.046). This minimum was  observed at the 
same phase, but  deeper than the optical minimum observed in 1993 (Panov,
Altmann \& Seggewiss 2000). By 2001 July, the optical light almost reached 
the pre-eclipse level. They interpreted this eclipse-like phenomenon as
due to the carbon dust envelope, the formation of which was triggered at 
the periastron passage. The difference in the rising part of the eclipse light
curves of 1993 and 2001 could be due to the differences in clumping in the 
expanding dust shell from cycle to cycle.  Aperture masking interferometric 
observations by Monnier, Tuthill \& Danchi (2002) using the Keck Telescope 
showed the first direct images of dust in WR 140. Comparison 
of their pre-periastron observation in 1999 June and post-periastron 
observation in 2001 July at 2.2 $\mu$m shows expanding clumpy dust.
They resolved five clumps in the shell,  expanding with velocities 
0.46--1.29 milliarcsec. day$^{-1}$.

\begin{table}
\begin{centering}
\caption{Log of Spectroscopic Observations}
\label{obslog}
\begin{tabular}{@{}lllll}
\hline
\multicolumn{5}{c}{\underline{UKIRT + CGS4 (150 lines/mm grating, R=4700)}}\\
UT Date   	&JD		&Bands/			&Phase  &$d/a$	    \\
{\tiny yyyymmdd.dddd}  	&2450000+	&$\lambda (\mu m)$	&	&		\\
\hline
20001013.3495  &1830.85	&1.07,1.1 		&0.9605	&0.5092     \\
20001225.1921  &1903.69	&1.083			&0.9856	&0.2506     \\
20001226.2004  &1904.7	&1.083			&0.9860 &0.2465     \\
20010318.654   &1987.15	&1.083			&1.0144 &0.2507     \\
20010331.6208  &2000.12	&1.083 			&1.0189 &0.3018     \\
\hline

\multicolumn{5}{c}{\underline{UKIRT + CGS4 (40 lines/mm grating)}} \\
\multicolumn{5}{c}{\underline{R=800 in J band, R=400 in H and K bands}}\\
\hline
20010331.6618  &2000.16	&J,H,K	     &0.0189	&0.3020	\\
20010428.6164  &2028.12	&J,K	     &0.0285	&0.4045	\\
20010521.5254  &2051.03	&J,H,K	     &0.0364    &0.4810	\\
20010605.4691  &2065.97	&J,K	     &0.0416    &0.5277 \\
20010613.5553  &2074.06	&J           &0.0444    &0.5521 \\
20010704.6004  &2095.1	&J,H,K       &0.0516    &0.6129 \\
20010812.2456  &2133.75	&J,H,K       &0.0650    &0.7156 \\
20010909.3325  &2161.83	&J,H,K       &0.0747    &0.7841 \\
20011006.2138  &2188.71	&J,H,K       &0.0839    &0.8457 \\
20011122.2333  &2235.73	&J,K         &0.1001    &0.9452 \\
20011226.1985  &2269.7	&J,K         &0.1119    &1.0115 \\
20020403.6638  &2368.16	&J,H,K       &0.1458    &1.1816 \\
20020630.3636  &2455.86	&J,H,K	     &0.1761    &1.3104 \\
20020717.5517  &2473.05	&J,H,K       &0.1820    &1.3335 \\
\hline
\multicolumn{5}{c}{\underline{UKIRT + UIST (IJ, HK \& short$\_$J grisms)}} \\
\multicolumn{5}{c}{\underline{R=3000-short$\_$J, R=950-IJ \& R=800-1000-HK grism}}\\
20030524.5496  &2784.05	&short$\_$J  &0.2893    &1.6542 \\
20030524.5640  &2784.06	&IJ	     &0.2893	&1.6542	\\
20030524.5740  &2784.07	&HK          &0.2893    &1.6542 \\
\hline
\multicolumn{5}{c}{\underline{Mt. Abu (R=1000)}}\\
19980508.9457  &0942.45	&JHK	     &0.6540	&1.7620 \\
20001221.575   &1900.08	&JH          &0.9844    &0.2650	\\
20001222.5658  &1901.07	&J           &0.9847	&0.2610 \\
20001223.5958  &1902.10 	&J           &0.9851    &0.2570 \\
20010102.5861  &1912.09	&J           &0.9885    &0.2168 \\
20010103.6090  &1913.11	&J           &0.9889    &0.2127 \\
20010326.9995  &1995.50	&J           &0.0173    &0.2836 \\
20010328.9988  &1997.50	&J           &0.0180    &0.2914 \\
20010402.9167  &2002.42	&J           &0.0197    &0.3106 \\
20010403.8743  &2003.37	&J           &0.0200    &0.3145 \\
\hline
\end{tabular}
\end{centering}
\end{table}

The programme of near-infrared spectroscopy described in the present 
paper was undertaken to search for changes in the infrared line spectrum, 
particularly the profile of the He\,{\sc i} line at 1.083 $\mu$m, around 
the time of maximum colliding wind activity. Previous 1-micron and $JHK$ 
spectroscopy, e.g. Vreux, Andrillat \& Bi\'emont (1990), Eenens, Williams 
\& Wade (1991), and observations of the 1.083-$\mu$m He\,{\sc i} line
profile (Eenens \& Williams 1994), were made at phases far from periastron. 
In each observation, the 1.083-$\mu$m emission-line profile showed a flat 
top, characteristic of formation in the asymptotic region of the Wolf-Rayet 
wind, with no evidence of a sub-peak which could be formed in a wind 
collision region (Stevens \& Howarth 1999). Nevertheless, given the 
very high eccentricity of the binary system and short duration of 
maximum colliding-wind activity such as dust formation, we considered it 
worth reobserving the 1.083-$\mu$m profile as close to the periastron as 
possible. 

\section{Observations}

We observed near-IR spectra of WR 140 using the 3.8-m United Kingdom
Infrared Telescope (UKIRT) on Mauna Kea, Hawaii and the 1.2-m Mt. Abu 
Infrared Telescope, India.  The UKIRT observations were made using the 
Cooled Grating Spectrometer (CGS4) and the UKIRT 1--5 micron Imager 
Spectrometer (UIST) while the Mt. Abu observations were made using an 
LN2 cooled NICMOS array imager and spectrometer.  We observed WR 140 
from 2000 October to 2003 May, covering the periastron passage expected 
in 2001 February. No observations were possible during January and 
February since WR 140 was not available at night.  From 2000 October 
to 2001 March, the He\,{\sc i} line at 1.083 $\mu$m was observed
using CGS4 and 150 lines/mm grating at a spectral resolution of
$\sim$4700, after which this grating was not available for observations.
In 2003 May, this line was again observed using  UIST with the short J grism, 
which gives a spectral resolution of 3000 with a 2-pixel slit.  From
2001 March, observations were performed in  $J$, $H$ and $K$ bands using 
CGS4 and the 40 lines/mm grating, which gives a resolution of 800 in the $J$ 
band in the second order and 400 in the $H$ and $K$ bands in the first order.
Observations using the Mt. Abu spectrometer were done in the $JHK$ bands
\begin{figure}
\vspace*{220pt}
\includegraphics{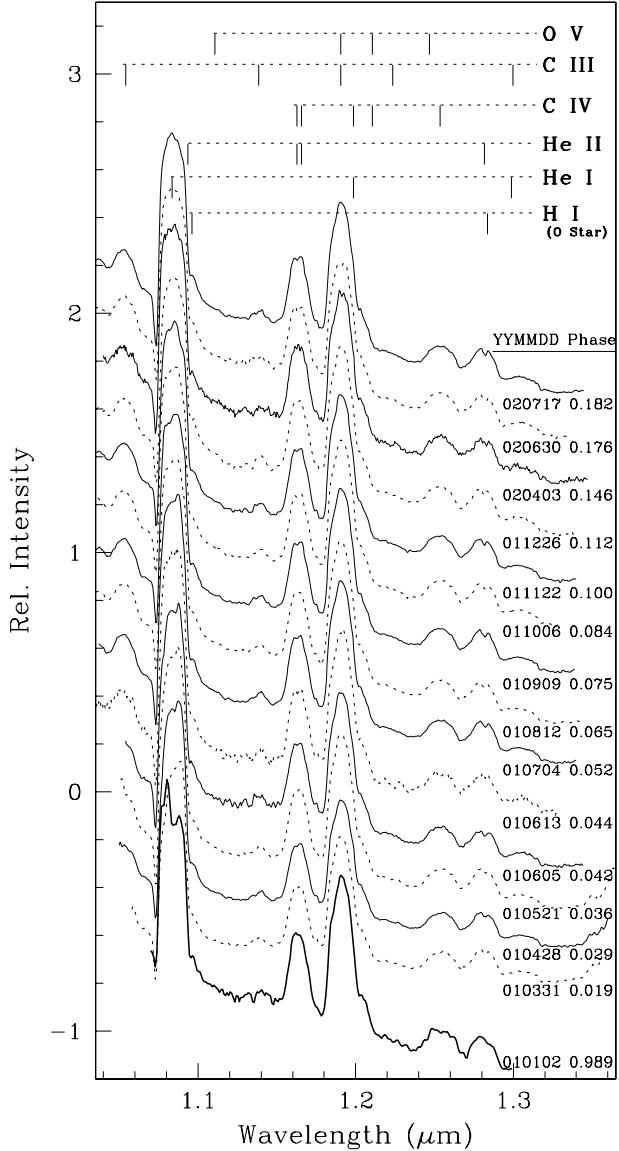}
\vspace*{230pt}
\caption{ $J$ band spectra of WR 140 observed using UKIRT and CGS4.  Spectra 
are labelled with the UTdate and phase of observation.  The $J$ spectrum 
observed using the Mt. Abu Telescope on 2001 January 2 is shown at the 
bottom of the montage for comparison. All spectra are plotted on the same 
scale with vertical shifts for clarity.}
\label{wr140jmk}
\end{figure}
\begin{figure}
\vspace*{220pt}
\includegraphics{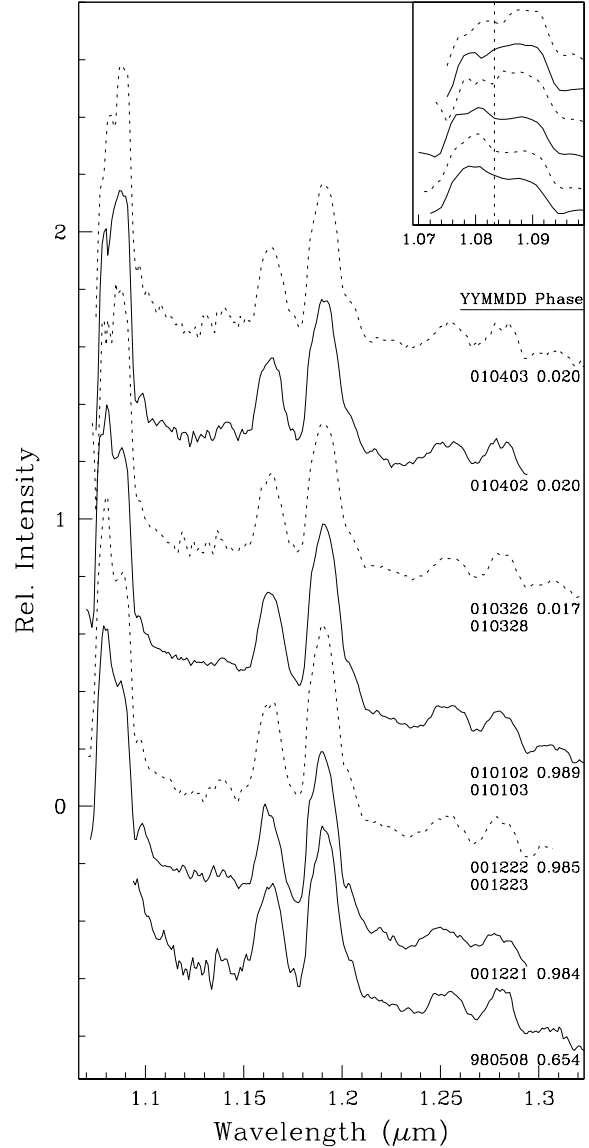}
\vspace*{230pt}
\caption{$J$-band spectra of WR 140 observed using the 1.2m Mt. Abu IR 
telescope and the NICMOS spectrometer with spectra observed on consecutive 
days averaged. The spectra are labelled with the UTdate and phase of observation. 
All spectra are plotted in the same scale with vertical shifts for clarity. 
The inset shows an expanded view of the evolution of the 1.083 $\mu$m
He\,{\sc i} line. The dashed vertical line shows the central wavelength of 
the line.}
\label{wr140jabu}
\end{figure}
at a resolution of $\sim$1000 in 1998 May and  from 2000 Dec. to
2001 April. Due to decreasing transmission of the blocking filter
towards shorter wavelengths, the S/N of the J spectra observed by
the Mt. Abu spectrometer decreases towards 1.083 $\mu$m. However,
the line profile variations of the He\,{\sc i} line are clearly seen.
Table {\ref{obslog}} gives details of the spectroscopic observations.
In the last column, $d$ refers to the separation between the O
star and the WC star at any epoch of observation and $a$ refers to the
semi-major axis of the relative orbit with the WC star at one of
the focii.

\begin{figure*}
\vspace*{200pt}
\includegraphics{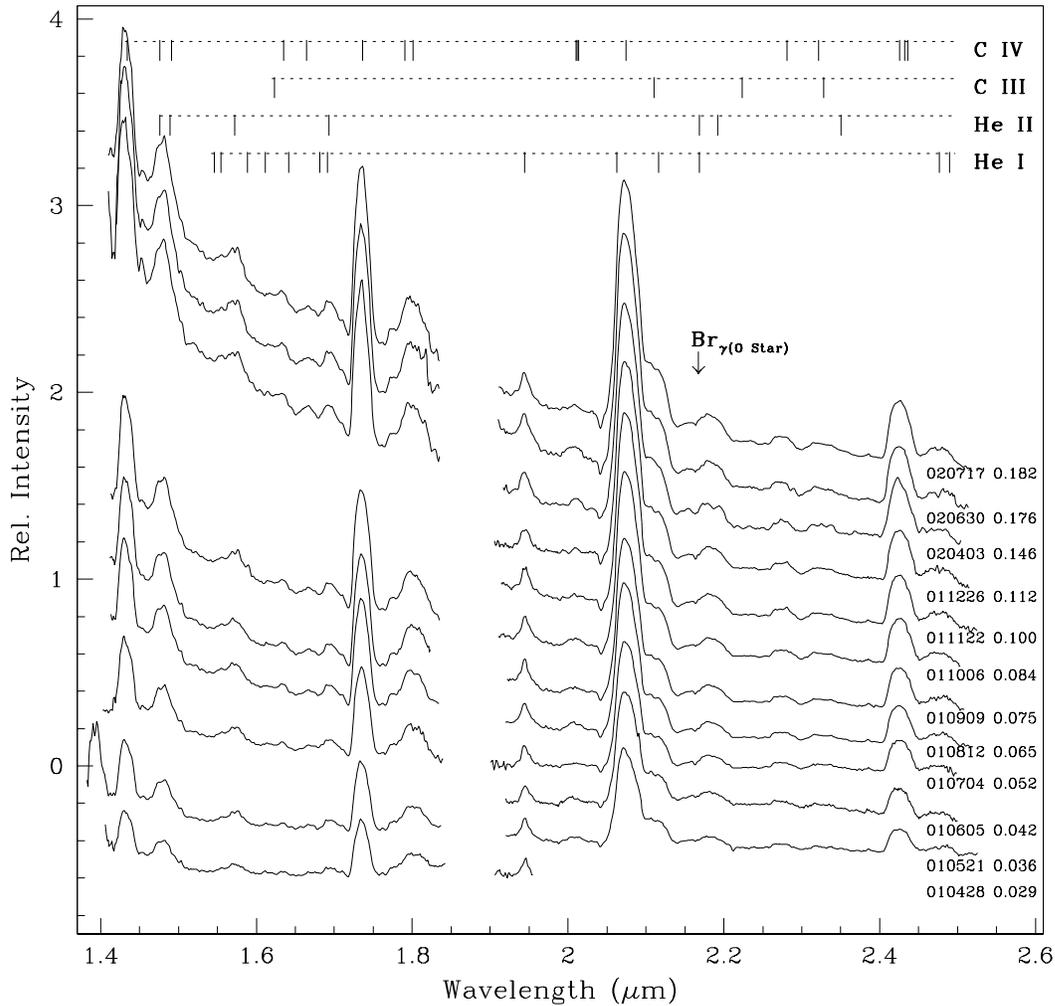}
\vspace*{200pt}
\caption{$H$ and $K$-band spectra of WR 140 observed using UKIRT and CGS4. 
Spectra are labelled with their UT Date of observation and the orbital phase}
\label{wr140hk}
\end{figure*}


\section{The $JHK$-band spectra}

\begin{figure*}
\vspace*{100pt}
\includegraphics{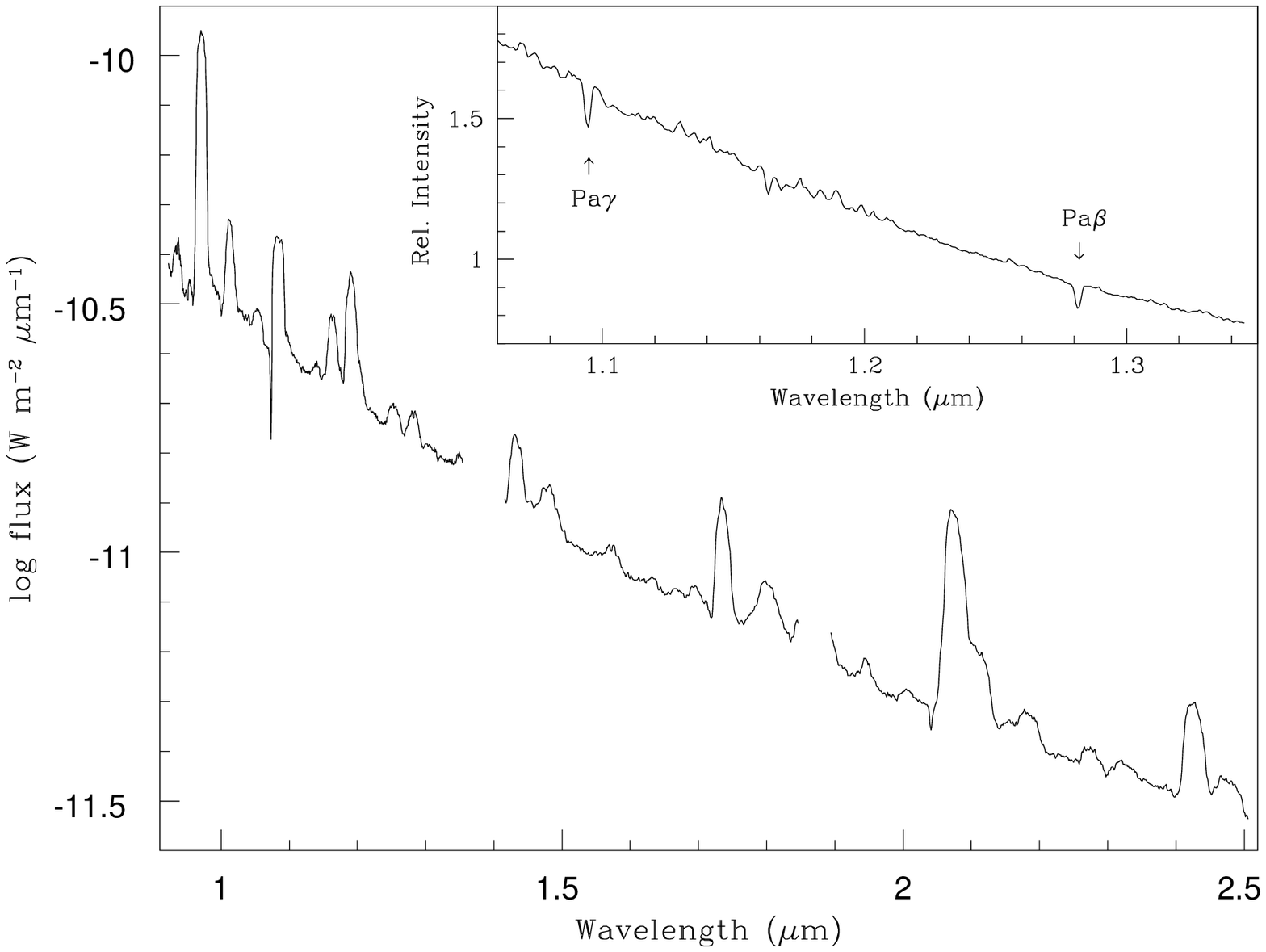}
\vspace*{220pt}
\caption{Flux-calibrated $JHK$ spectra of WR 140 observed using UKIRT and UIST
on 2003 May 24.
The box shows the $J$ band spectrum of the O star BS 6736 (9 Sgr) observed on 
2001 June 5. The conspicuous absorption lines in it are Pa$\gamma$ and 
Pa$\beta$ at 1.094 and 1.282 $\mu$m respectively.}
\label{wr140zjhk}
\end{figure*}

Figs. {\ref{wr140jmk}} and {\ref{wr140jabu}} show the $J$-band spectra 
observed from Mauna Kea and Mt. Abu respectively.  One of the Mt. Abu 
spectra (2001 Jan 2) is also shown in Fig. {\ref{wr140jmk}} along with 
the $J$ spectra of later epochs from Mauna Kea for comparison of the 
continuum. Fig. {\ref{wr140hk}} shows the $H$ and $K$ spectra. 
The gap in the  spectra is the region where the telluric absorption 
makes the atmospheric transmission poor.  Each spectrum is labelled with 
the date of observation and orbital phase of the binary system. 
Fig. {\ref{wr140zjhk}} shows the observed, flux-calibrated spectra
observed on 2003 May 24 using UKIRT and UIST. The data were flux calibrated
using photometry from Mt. Abu Observatory on 2003 June 5 and 12, which 
gave average magnitudes of $J=5.61 \pm0.07$ and $H=5.34 \pm0.07$ 
respectively.  Spectra of 2003 May 24 are dereddened assuming E(B-V) = 0.6 
(Eenens \& Williams 1992) and the equivalent widths of the spectral lines
are determined fitting Gaussian profiles.  Multiple Gaussians are fitted 
when the lines are blended.  Table {\ref{lineid}} lists the spectral lines 
identified and their fluxes and equivalent widths, estimated from the
de-reddened spectra. 1$\sigma$ errors are given in brackets against the
estimated values.  The main source of errors is from the definition of the
continuum.  The error estimates are done from multiple attempts of the fits.
Line identifications are done in comparison with 
the near IR spectra of WR 140 and the identifications of
Eenens \& Williams (1992) and from the atomic line list of Peter van 
Hoof{\footnote{van Hoof, P. A. M.,   http://star.pst.qub.ac.uk/$^\sim$pvh/}}. 
For the 1.083 $\mu$m He\,{\sc i} line, since the line was flat topped, 
multiple lines were fitted to estimate the line flux and EW. A
comparison of the line fluxes with those of Eenens \& Williams (1992) 
(spectra taken in 1988 July, phase=0.41) shows that the line fluxes 
of most of the prominent lines match well, implying that the wind 
properties of the system are very much repeatable.

\begin{table*}
\begin{centering}
\caption{Line identifications, line fluxes, Equivalent Widths (EW) and FWHM from the spectrum
observed on 2003 May 24 and de-reddened assuming E(B-V)=0.6.  1$\sigma$ errors are given in brackets.}
\label{lineid}
\begin{tabular}{@{}llllll}
\hline
$\lambda$& Main &Other probable&Flux	&EW & FWHM \\[-1mm]
$(\mu$m) &contributor &contributors &(10$^{-15}$W/m$^{2}$)&(\AA) & km/s \\[-1mm]
\hline
0.933   &C {\sc iii}    &0.9347(He {\sc ii}), 0.936,0.941(C {\sc iii})	&153 (5.7)      &-20.3 (0.9)   	&3160 (67)   \\[-0.8mm]
	&		&0.934(C {\sc iv})				&		&		&	\\[1mm]
0.9529  &He {\sc i}	&0.954(O {\sc iii})				&26 (2)		&-3.7 (0.3)	&	\\[1mm]
0.972	&C {\sc iii}	&0.975,0.965 (C {\sc iii}), 0.977(O {\sc v})	&2431 (5)	&-371 (2.1)	&	\\[-0.8mm]
0.987   &C {\sc iv}     &0.984(O {\sc v})      				&122 (13)	&-19.2 (2.2)  	&        \\[1mm]
1.0126	&He {\sc ii}	&1.003(He {\sc i}, C {\sc iv}), 1.011-1.023(C {\sc iii})&380 (3)&-62 (0.7)	&3320 (30)	\\[1mm]
1.055	&C {\sc iv}     &1.055(C {\sc iii}), 1.053(C {\sc iv})		&82 (2.2)	&-16 (0.5)	&3412 (45)    \\[1mm]
1.083   &He {\sc i}     &1.078(O {\sc v}), 1.091(He {\sc i})		&558 (3)	&-121 (1) 	&        \\[1mm]
1.094   &He {\sc ii}    &       					&44 (2.5) 	&-10 (1)	&        \\[-0.8mm]
1.1 	&C {\sc iv}     & 						&		& 		&        \\[1mm]
1.139 	&C {\sc iv}   	&		       				&30 (2.2)	&-7.5 (0.6)    	&2800 (120)    \\[1mm]
1.163 	&He {\sc ii}    &1.163(C {\sc iv}), 1.164(C {\sc iii})		&220 (1)	&-60 (1)  	&        \\[-0.8mm]
1.165   &C {\sc iii}    &1.168(He {\sc ii})   				&		&       	&        \\[1mm]
1.191 	&C {\sc iv}     &1.188(C {\sc iv}), 1.191(O {\sc v})		&424 (2.5)	&-124 (1)    	&        \\[-0.8mm]
1.199   &C {\sc iii}    &1.197(He {\sc i})				&		&       	&        \\[1mm]
1.205	&O {\sc v}      &1.211(C {\sc iii})				&47 (3)		&-14 (1.5)  	&        \\[1mm]
1.226 	&C {\sc iv}     &               				&7.3 (1)    	&-2.3 (0.3)    	&        \\[1mm]
1.247 	&O {\sc v}      &						&66 (1.4)	&-23 (1)  	&        \\[-0.8mm]
1.255 	&C {\sc iii}    &1.253(He {\sc i}), 1.256(C {\sc iii})		&		&       	&        \\[-0.8mm]
	&               &1.258,1.261(C {\sc iii}), 1.255(O {\sc v})	&		&       	&        \\[1mm]
1.282 	&He {\sc ii}    &1.279,1.285(He {\sc i}), 1.277(C {\sc iv})	&70 (1.3)	&-27 (0.6)     	&3600 (30)   \\[1mm]
1.298 	&C {\sc iv}     &1.307(C {\sc iv}), 1.297(He {\sc i})		&21.8 (2.3)	&-9 (1)    	&        \\[1mm]
1.435 	&C {\sc iv}     &1.434(C {\sc iv})     				&140 (3)	&-78 (1.8)    	&3637 (85)    \\[1mm]
1.454 	&C {\sc iii}    &               				&22 (5)		&-13 (3)    	&        \\[1mm]
1.476 	&He {\sc ii}    &1.473(C {\sc iii}), 1.47(O {\sc v}) 		&88 (2.1)	&-55 (2)     	&        \\[-0.8mm]
1.491 	&C {\sc iv}     &1.489(He {\sc ii})     			&       	&       	&	 \\[1mm]
1.547 	&He {\sc i}     &1.547,1.551(O {\sc iv})			&5.5 (0.7)	&-4 (0.6)  	&        \\[-0.8mm]
1.552 	&He {\sc i}     &               				&       	&       	&	 \\[1mm]
1.572 	&He {\sc ii}    &               				&27 (0.7)	&-21 (0.5)    	&        \\[-0.8mm]
1.588 	&He {\sc i}     &1.58-1.59(O {\sc v})  				&       	&		&        \\[1mm]
1.635 	&C {\sc ii}     &1.635(O {\sc v})       			&8.5 (0.8)	&-7.5 (0.7)     &2700 (225)    \\[-0.8mm]
1.641 	&He {\sc i}     &               				&     		&		&        \\[1mm]
1.664 	&C {\sc iv}     &               				&10.7 (0.7)	&-9.8 (0.7)    	&	 \\[1mm]
1.693 	&He {\sc ii}    &1.712(C {\sc iv}), 1.699(O {\sc v}) 		&25 (1)		&-24.8 (1)   	&	 \\[-0.8mm]
1.701 	&He {\sc i}     &               				&		&		&	 \\[1mm]
1.736 	&C {\sc iv}     &1.736,1.737(C {\sc iv})			&148 (0.5)	&-152 (0.5)    	&3240 (25)    \\[1mm]
1.785 	&C {\sc ii}     &1.790(C {\sc iv})    				&83 (3)		&-92 (3)    	&        \\[-0.8mm]
1.801 	&C {\sc iv}     &1.814(He {\sc i}), 1.819(C {\sc iv}), 1.820(C {\sc ii})&     	&		&       \\[1mm]
1.944 	&He {\sc i}     &               				&13.5 (0.7)	&-19.5 (1)   	&2530 (85)    \\[1mm]
2.012 	&C {\sc iv}     &2.010(C {\sc iv})    				&6.0 (0.6)	&-9.5 (0.95)	&        \\[1mm]
2.059 	&He {\sc i}     &               				&284 (3.4)	&-493 (7)    	&        \\[-0.8mm]
2.071 	&C {\sc iv}     &2.080(C {\sc iv})    				&       	&       	&	 \\[1mm]
2.107 	&C {\sc iv}   	&2.108(C {\sc iii})   				&71 (1.2)	&-130 (2.5)   	&        \\[-0.8mm]
2.113   &He {\sc i}     &       					&       	&		&        \\[-0.8mm]
2.122   &C {\sc iii}    &2.139(C {\sc iii})  				&       	&		&   	 \\[1mm]
2.150   &He\,{\sc i}	&						&7.3 (0.8)	&-13.9 (1.4)	&	 \\[-0.8mm]
2.165 	&He {\sc i}, He {\sc ii}&2.161(He {\sc i}, C {\sc ii}), 2.274(C {\sc ii})&   	&     		&	 \\[1mm]
2.189	&He {\sc ii}    &          					&24 (1.4)    	&-47 (3)   	&        \\[1mm]
2.278   &C {\sc iv}     &               				&10 (1)		&-23.4 (2.3)   	&3080 (155)    \\[1mm]
2.318 	&C {\sc iv}     &2.314(He {\sc ii})   				&13 (0.7)	&-31 (1.6)	&        \\[-0.8mm]
2.325 	&C {\sc iii}    &2.328,2.371(C {\sc iv})			&       	&		&        \\[-0.8mm]
2.347 	&He {\sc ii}	&       					&     		&		&        \\[1mm]
2.423 	&C {\sc iv}     &       					&65 (0.7)   	&-175 (1.9)	&        \\[-0.8mm]
2.429   &C {\sc iv}     &2.426(C {\sc iv})				&       	&		&        \\[-0.8mm]
2.433   &C {\sc iv}     &2.432(O {\sc v})   				&       	&		&        \\[1mm]
2.473   &He {\sc i}     &2.472(He {\sc i}), 2.470(O {\sc v})		&22 (0.8)   	&-62 (2.6)	&	 \\[-0.8mm]
2.486   &He {\sc i}     &						&		&		&	 \\[1mm]
\hline
\end{tabular}
\end{centering}
\end{table*}

\begin{table*}
\begin{centering}
\caption{Equivalent widths of emission lines}
\label{ewevoln}
\begin{tabular}{@{}lllllllll}
\hline
Date     &Phase 	&1.165 &1.255 &1.282 &1.435 &1.736 &1.944 &2.43  \\
       	 &		&EW    &EW    &EW    &EW    &EW    &EW    &EW    \\
\hline
19880715 &0.4176        &-55   &-29.44&-30.3 &-78.6 &	   &      &      \\
19980508 &0.6541	&-66.8 &-27.6 &-38.  &      &-163.0&      &      \\
20001221 &0.9844	&-57.0 &-35   &-33   &      &      &      &      \\
20001223 &0.9851	&-68.9 &-24.0 &-28.0 &      &      &      &      \\
20010102 &0.9885	&-62.4 &-35.4 &-30.3 &      &      &      &      \\
20010103 &0.9889	&-55.0 &-30.1 &-36.9 &      &      &      &      \\
20010326 &0.0173	&-50.1 &-21.4 &-24.4 &      &      &      &      \\
20010328 &0.0179	&-49.8 &-20.7 &-26.2 &      &      &      &      \\
20010331 &0.0189	&-47.8 &-19.5 &-22.0 &      &      &      &      \\
20010402 &0.0197	&-55.0 &-26.6 &-29.0 &      &      &      &      \\
20010403 &0.0200	&-56.2 &-24.4 &-26.7 &      &      &      &      \\
20010428 &0.0285	&-49.4 &-20.6 &-22.4 &-37.9 &-60.7 &      &      \\
20010521 &0.0364	&-58.2 &-22.7 &-26.8 &-44.5 &-73.0 &-13.4 &-40.2 \\
20010605 &0.0416	&-57.2 &-21.8 &-24.0 &      &      &-12   &-47.6 \\
20010613 &0.0444	&-50.0 &-25.2 &-28.6 &      &      &      &      \\
20010704 &0.0516	&-59.9 &-23.8 &-27.7 &-51.0 &-92.0 &-14.0 &-50.9 \\
20010812 &0.0650	&-56.3 &-24.2 &-27.1 &-51.9 &-104.0&-15.8 &-65.6 \\
20010909 &0.0747	&-57.1 &-24.0 &-28.0 &-55.8 &-110.0&-16.4 &-69.1 \\
20011006 &0.0839	&-59.9 &-23.9 &-28.0 &-59.7 &-112.5&-17.9 &-79.0 \\
20011122 &0.1001	&-58.2 &-24.3 &-27.8 &      &      &-15.6 &-90.0 \\
20011226 &0.1119	&-61.4 &-25.1 &-29.6 &      &      &-15.8 &-94.3 \\
20020403 &0.1458	&-58.5 &-25.3 &-31.2 &-62   &-125.6&-21.6 &-108.5\\
20020630 &0.1761	&-56.7 &-26.6 &-30.3 &-65   &-131.8&-21   &-126.5\\
20020717 &0.1820	&-58.4 &-27.1 &-31.9 &-64.7 &-135.2&-20.3 &-131.0\\
20030524 &0.2893	&-63.1 &-23   &-26.1 &-76.7 &-148.8&-24.2 &-170.6\\
\hline
&&&\\
\end{tabular}
\end{centering}
\end{table*}

\begin{figure}
\vspace*{180pt}
\includegraphics{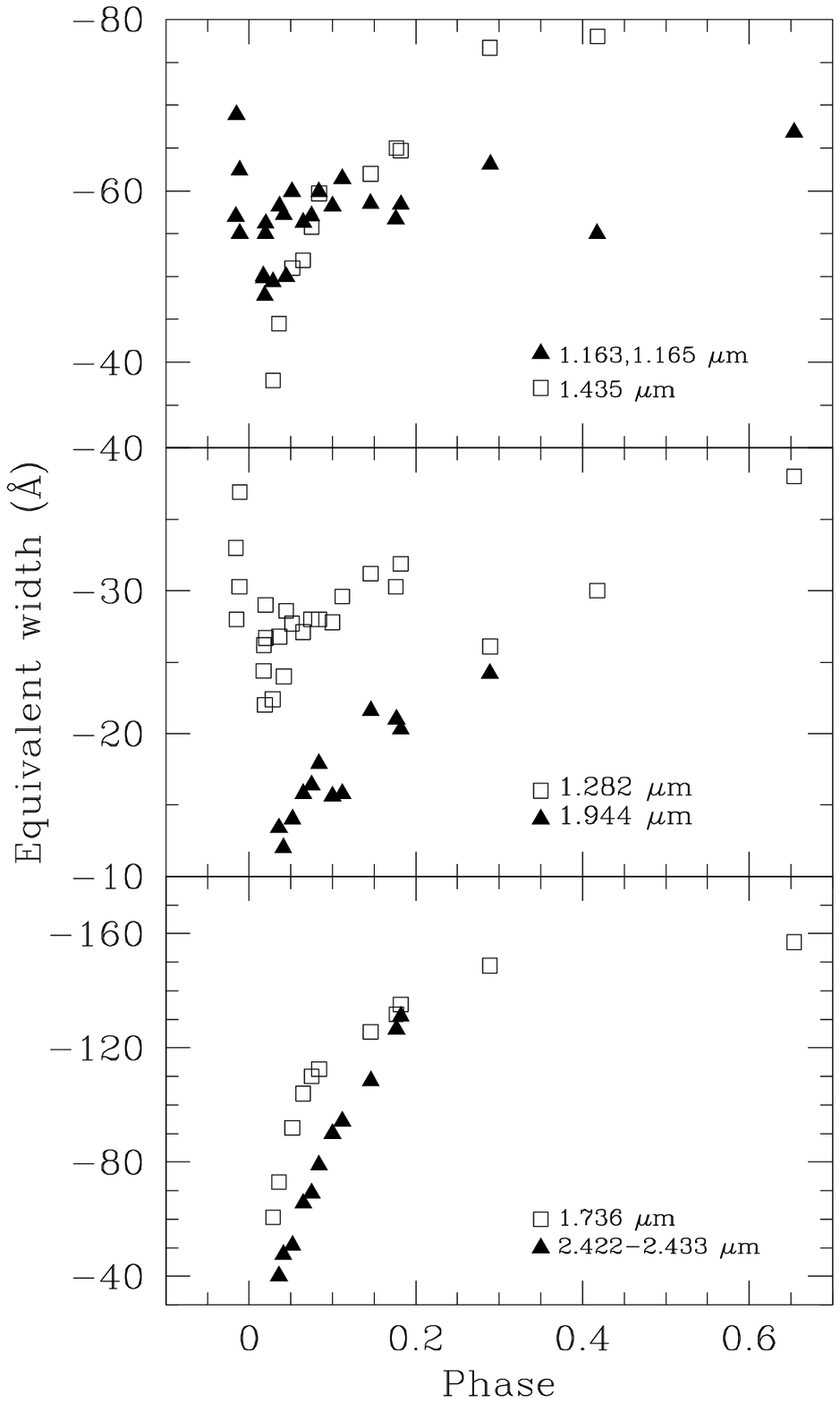}
\vspace*{200pt}
\caption{The variation of equivalent width of different emission lines
due to dust formation and cooling}
\label{wr140ew}
\end{figure}

As a measure of the influence of dust emission on the $JHK$ 
spectra, we give the equivalent widths of prominent emission lines 
estimated from the observed spectra
in Table \ref{ewevoln}. The 1$\sigma$ errors are $\sim$ $6-7 \%$ 
of the EW for the He {\sc i} line at 1.945 $\mu$m. For the
rest of the lines they are $\leq$ $5 \%$.
Assuming that the fluxes within the 
emission lines did not change, it is evident from the EWs of those 
in the 1.25--1.28 $\mu$m region that dust formation began between 2001 
January 3 and March 26 (orbital phases 0.989 and 0.017).
At the time of the 2001 March 26 spectrum,
light curves from previous cycles predict contributions
(relative to the total flux) by dust emission of 22$\%$ in J,
55$\%$ in H and 76$\%$ in K, i.e. dust-to-wind ratios 0.28, 1.2
and 3.2. These are a strong function of wavelength within
the bands, e.g. rising from 0.24 at 1.16 micron to 0.32
at 1.28 micron. This is consistent with infrared light curves from 
previous cycles (W90), which predict dust emission with a steeply
rising spectrum having at phase 0.017 a flux ratio of 0.28
to the stellar wind in the $J$ band. The EWs of the 1.166-$\mu$m 
feature are harder to interpret, partly bcause of increased scatter 
on their measured strengths and partly because dust emission at this 
wavelength is about two-thirds that at 1.25 $\mu$m (from the model 
fit to the photometry at this phase in W90). 
Also, as can be seen in Fig. \ref{wr140jmk}, the slope of the continuum  
of the $J$ spectrum taken in 2001 March 31 is much flatter than that 
taken on January 2. By the time of the 2003 May observation,
when the light curves predict that the dust emission
contribution to the total flux had fallen to 10 per cent in
the $K$ band and below 5 per cent in $J$ and $H$ bands owing
to the cooling of the dust, the slope in the $J$ region had
returned to that before dust formation.
The $H$ and $K$ spectra show the change in continuum caused by variable 
amount of dilution due to emission from the dust. The $HK$ region shows 
a nearly flat spectrum during 2001 April when the dust emission was at 
maximum at these wavelengths, and the spectra became steeper when the 
dust became cooler.  
The equivalent widths (Table \ref{ewevoln}, which includes EWs estimated 
from the $J$ spectrum of Eenens, Williams \& Wade (1991), observed in 
1988 with UKIRT and CGS2) of six of these lines are plotted as a function 
of orbital phase in Fig. {\ref{wr140ew}}.  This shows the effect of the 
dilution due to the thermal emission from the dust and the dust cooling.
There is a phase lag in the rise of equivalent width with wavelength.
This is due to the cooling of the dust and peak emission from the dust 
shifting to higher wavelengths due to cooling.

\section{The 1.083-$\mu$m He\,{\sc i} line profile}
\subsection{The absorption component of the profile}

\begin{figure*}
\vspace{180pt}
\includegraphics{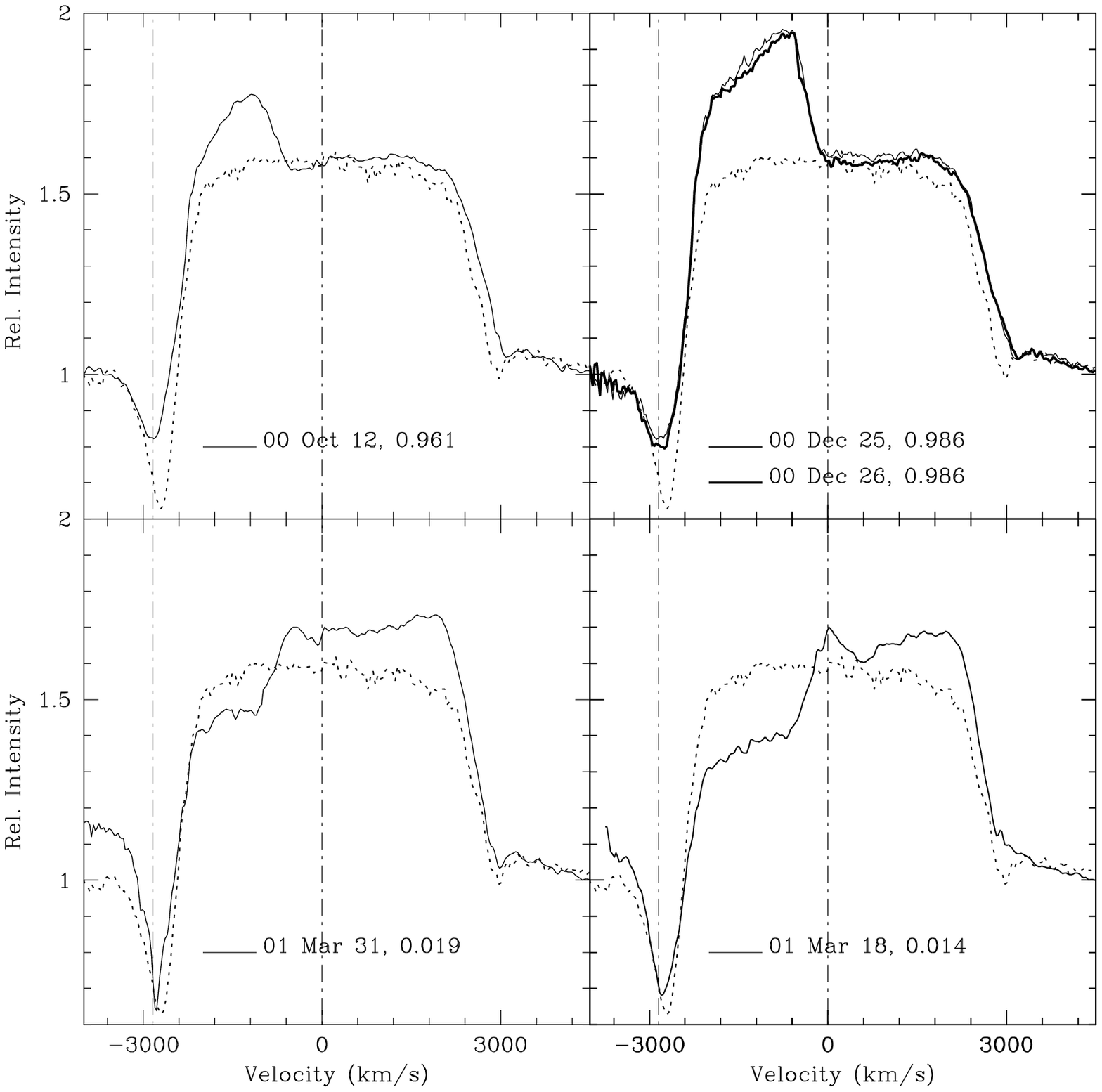}
\vspace{180pt}
\caption{High resolution spectra of the 1.083 $\mu$m He\,{\sc i} line observed 
with UKIRT, corrected for the radial velocity of the WC star, labelled with UT 
date of observation and orbital phase.  High resolution spectrum observed on
2003 May 24 is overplotted (dotted) on all the spectra for comparison. The 
dashed vertical lines are at velocities 0 and -2840 km s$^{-1}$.  Variations 
in both the emission component and the absorption component of the P-Cygni 
profile are evident.}
\label{wr1401083}
\end{figure*}

The UKIRT high-resolution spectra covering the 
He\,{\sc i} line (2s$^{3}$S -- 2p$^{3}$P) at 1.083 $\mu$m, corrected 
for the line-of-sight velocity of the WC star at the time of observation, 
are shown in Fig. {\ref{wr1401083}}.
This line, and its singlet counterpart at 2.058 $\mu$m, show P Cygni  
absorption profiles from which wind terminal velocities $v_{\infty}$ of 
--2650 km s$^{-1}$ (Lambert \& Hinkle 1984), --2860 km s$^{-1}$ (Williams 
\& Eenens 1989) and --2900 km s$^{-1}$ (Eenens \& Williams 1994) have been 
measured. 
These velocities are in good agreement with that (--2900 km s$^{-1}$) 
derived from the violet edge of the saturated part of the UV C\,{\sc iv} 
P Cygni line profile, $v_{black}$, by Prinja, Barlow \& Howarth (1990), 
considered to be a better measure of the terminal velocity than the extreme 
violet edge of the absorption component, $v_{edge} = -3200$ km s$^{-1}$. 

From our high-resolution spectra of the 1.083-$\mu$m line, we measured 
$v_{edge}$, the velocity at the blue edge of this absorption profile where 
the absorption merges with the continuum, and $v_{abs}$, the velocity at 
the maximum of the P Cyg absorption with respect to the line centre.  
Two vertical lines are drawn in Fig. {\ref{wr1401083}}, at the line centre 
and at v = 2840 km s$^{-1}$, which is the $v_{abs}$ on 2000 December 25/26, 
before the periastron passage.  Table {\ref{velocity}} lists the $v_{edge}$ 
and $v_{abs}$ measured from the high resolution spectra. The errors in these 
velocities are within $\pm$ 30 km s$^{-1}$. The core velocities, $v_{abs}$, 
are consistent with the measurements of $v_{\infty}$ and show a smooth 
variation across the periastron passage, as can be seen well in Table 
{\ref{velocity}} and Fig.{\ref{wr1401083}}. The variations in $v_{abs}$ 
observed in the He\,{\sc i} line profile are unlikely to be caused by any 
He\,{\sc i} absorption profile in the wind of the O star. The inset in 
Fig. {\ref{wr140zjhk}} shows the $J$ spectrum of BS 6736 (9 Sgr), an 
O4 V((f)) star (Walborn 1972), observed with CGS4 in 2001 June. The spectrum
of 9 Sgr does not show any significant absorption at this wavelength.
We therefore assume that the observed He\,{\sc i} absorption occurs in
the WC wind material. 

As in the UV resonance line profiles, the values of $v_{edge}$ are greater  
than $v_{\infty}$, showing a maximum value of $\sim$ 3320 km s$^{-1}$ in 
2000 October and 2001 March and slightly less ($\sim$ 3250  km s$^{-1}$) 
in 2000 December and 2003 May. These velocities are close to that 
measured from the UV profiles ($\sim$ 3200 km s$^{-1}$) by 
Prinja et al. and most probably reflect observation of one or both of the 
stars through turbulent material near the wind-collision zone. 

\begin{figure*}
\vspace*{120pt}
\includegraphics{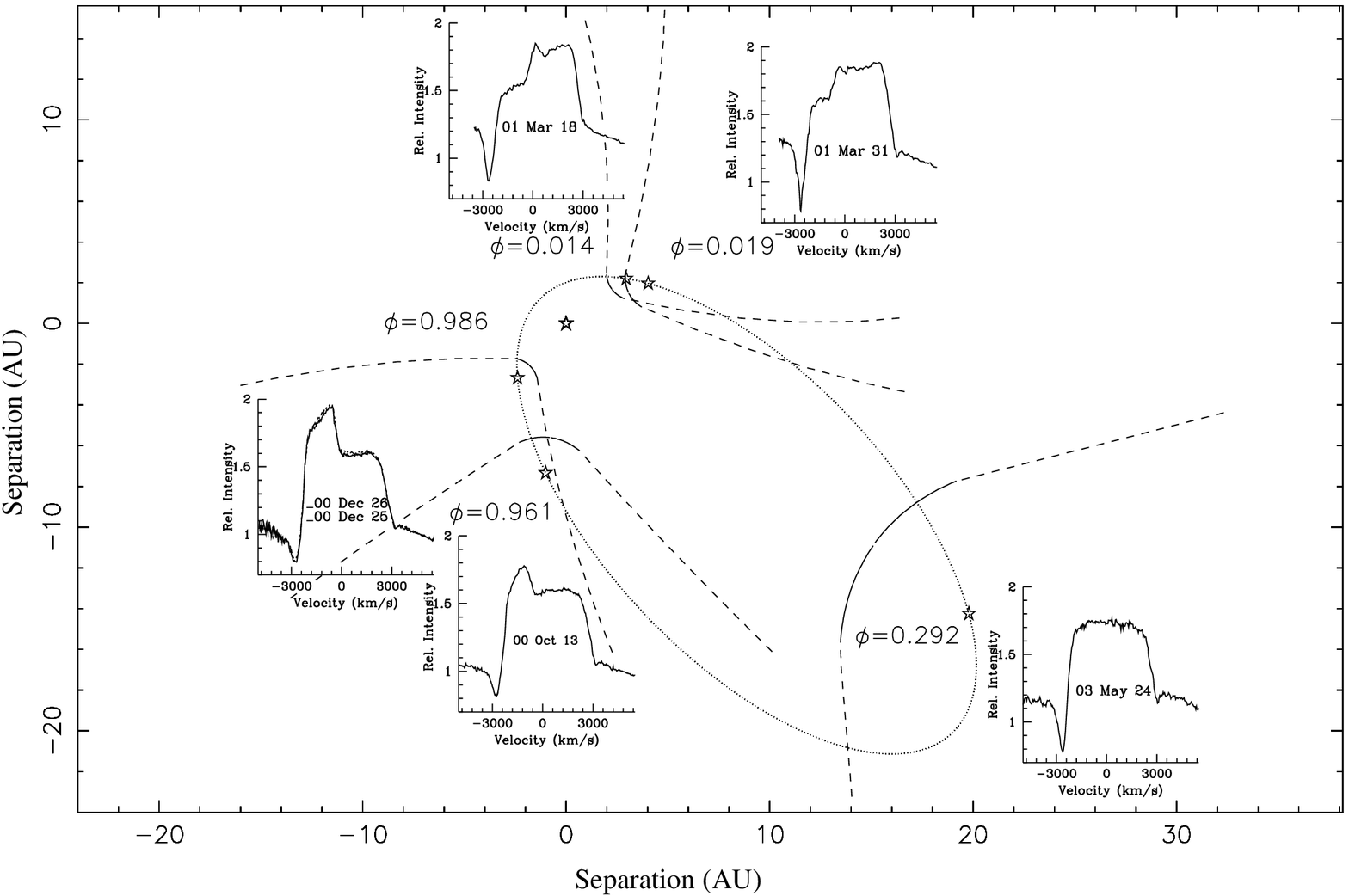}
\vspace*{190pt}
\caption{The relative orbit of the O star around the WR star, oriented so 
that the observer is at negative Y. 
The shape of the contact discontinuity is shown at the phases at which we 
observed high-resolution 1.083 $\mu$m He\,{\sc i} spectra. Superimposed on
the orbit are the 1.083 $\mu$m He\,{\sc i} line profiles, which are shown 
in more detail in Fig.{\ref{wr1401083}}}
\label{wr140orbit}
\end{figure*}

\begin{figure}
\vspace*{120pt}
\includegraphics{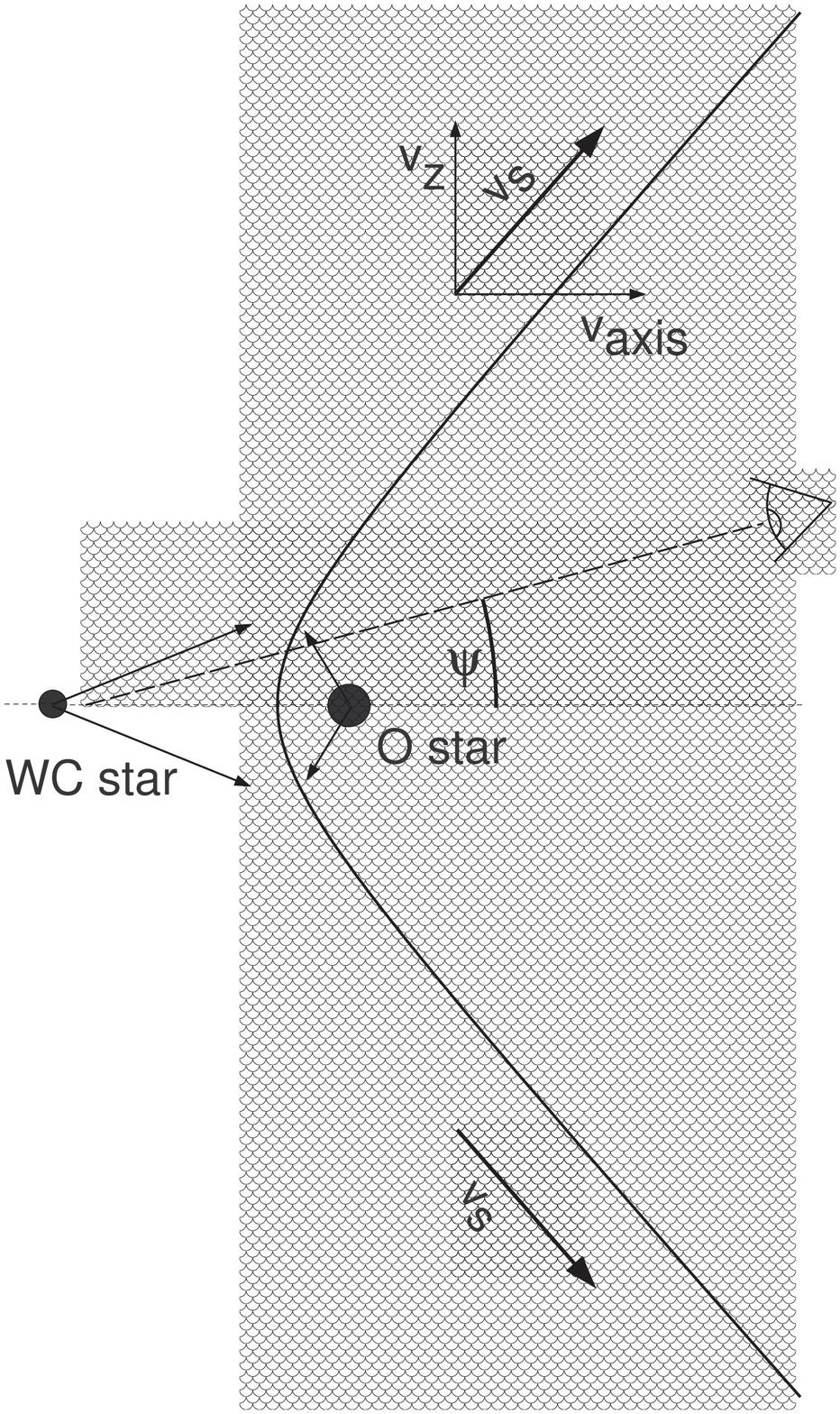}
\vspace*{150pt}
\caption{Sketch of the colliding-wind region between the WC and O-type stars
in WR 140 in the plane defined by the sightline to the observer and the axis
joining the stars. This plane is {\em not} perpendicular to the orbital plane
except at epochs of conjunction. The angle between the sightline and the
axis is $\psi$ and the wind interaction region is assumed to be symmetric
about this axis.
The hatched area represents compressed WC stellar wind material flowing
along the contact discontinuity between the two stellar winds.
At the time of the 2000 October observation, most of the material (the
fraction depending on the unknown orbital inclination) is flowing
towards the observer.}
\label{wr140oct}
\end{figure}

The depth of the P-Cygni absorption component varies considerably,
being greater in the 2001 March and 2003 May spectra than in the 2000
October and December spectra. We interpret this in terms
of variation of wind material in the sightline to the underlying stellar
continuum, which is provided by both the WC7 and O4-5 stars. The greatest
absorption was observed in the 2001 March spectra, when our sightline to
both stars passed through the He-rich WC stellar wind. This is illustrated
in Fig.~{\ref{wr140orbit}}, showing the configurations of the WR 140 system
at the epochs of the 1.083-$\mu$m spectra. This sketch is in the plane of
the orbit; the sightline is from the negative-Y direction and at an angle
to the plane of $(\pi/2 - i)$, where $i$ is the orbital inclination. On
the other hand, the absorption was very much smaller in the 2000 October
spectrum. At this phase, the system was nearly at conjunction and we
expect the sightline to the O4-5 star to pass through its
solar-composition wind (Fig.~{\ref{wr140orbit}}).
The configuration in the plane of the axis passing through the two
stars and the sightline at the time of the 2000 October observation is
shown in Fig.~{\ref{wr140oct}}. The relatively low He\,{\sc i} absorption
seen at this phase implies that the angle $\psi$ between the sightline and
the axis passing through the two stars is small enough that we observe both
stars mostly through the O-star wind, which we assume to provide significantly
less He\,{\sc i} absorption than the WC wind. Some He\,{\sc i} absorption
occurs in the WC wind before it hits the interaction region and some in the
latter, where turbulence probably accounts for the slightly higher $v_{edge}$
observed at this phase.
To interpret the variation of the He\,{\sc i} absorption, we must compare
the angle $\psi$, which depends on the orbital elements and phase, with
the geometry of the wind interaction region.

Both sketches show the contact discontinuities separating the winds. These
are defined as the surfaces where the ram pressures of the stellar winds
balance, and depend on the wind-momentum ratio
\begin{equation}
\eta = (\dot{M} v_{\infty})_{\rm O} / (\dot{M} v_{\infty})_{\rm WR}.
\label{etadef}
\end{equation}

\begin{table*}
\begin{centering}
\caption{Values of $v_{edge}$ and $v_{abs}$ for 1.083-$\mu$m spectra and the details of the sub-peak}
\label{velocity}
\begin{tabular}{@{}llllllllcl}
\hline
UTDate  &Phase  & $a/d$  &v$_{edge}$&v$_{abs}$&\multicolumn{4}{c}{sub-peak}	& PCyg  \\
\cline{6-9}
(yyyymmdd)  &       &	   &(km/s)&(km/s) &blue edge\footnotemark[1]	&red edge\footnotemark[1] &central vel.\footnotemark[1]& flux  	& abs.\\
	    &       &      &      &       &(km/s)   			&(km/s)		  	  &(km/s)                      &(W m$^{-2}$) 	&EW\\
\hline
20001013    & 0.960 &1.96 & 3316 & 2860  &-2105\footnotemark[2]  &-526	&-1284  &3.3 e-14\footnotemark[3]&2.4 	\\
20001225/6  & 0.986 &3.99/4.06 & 3254 & 2840  &-2167\footnotemark[2]  &-58	&-1056  &8.4 e-14\footnotemark[3]&2.0 	\\
20010318    & 0.014 &3.99 & 3327 & 2790  &-674		    	  &+2521&+1060  &$\sim$13.1 e-14	 &6.1	\\
20010331    & 0.019 &3.31 & 3315 & 2790  &-1080		  &+2674&+801   &$\sim$13.2 e-14	 &5.8 	\\
20030524    & 0.289 &0.61 & 3254 & 2720  &		          &	&	&0.          		 &5.2	\\
\hline
\multicolumn{9}{l}{\footnotemark[1] velocity with reference to the central wavelength of the WR HeI emission line}\\
\multicolumn{6}{l}{\footnotemark[2] upper limit - the real blue edge will be bluer than this} &\multicolumn{4}{l}{\footnotemark[3] lower limit}
\end{tabular}
\end{centering}
\end{table*}

The surface wraps around the star having the lower wind momentum, the
O star, on account of its lower mass-loss rate. It has a bow towards the
WC star and tends towards a `cone' on the other side of the O star.
Where the orbital motion is slow compared with the stellar wind, the
surface is symmetric about the axis joining the WC and O stars. In the
case of WR 140, this holds for most of the orbit, except near periastron,
when the orbital motion is great enough to cause the interaction region
to trail behind the axis at large distances from the stars
(Fig. {\ref{wr140orbit}}). Here, we are concerned with absorption in
the winds, which occurs close to the stars, and will ignore the
azimuthal `twisting' of the interaction regions.
Analytical formulae for the form of the contact discontinuity are given
by Eichler \& Usov (1993) and Cant\'o, Raga \& Wilkin (1996). For
comparison with the angle $\psi$, we use the opening semi-angle $\theta$
of the `cone' given by Eichler \& Usov (1993)

\begin{equation}
\theta \simeq 2.1(1-\frac{\eta ^{2/5}}{4})\eta ^{1/3}
\label{eta2theta}
\end{equation}
                                                                                               
The angle $\psi$ is given by

\begin{equation}
cos(\psi) = -sin(i)  sin(f+\omega),
\label{psidef}
\end{equation}

\noindent where $f$ is the true anomaly and $\omega$ the argument of
periastron. We used the value of $\omega$ determined by Marchenko et al.
(2003) and their orbital elements to calculate $f$ for each observation.
Only the inclination, $i$, is unknown.

Setia Gunawan et al. (2001) derived $i=38^{\circ}$ from the red-shifted
absorption components the C\,{\sc iv} and Si\,{\sc iv} resonance lines
observed in an IUE spectrum observed at phase 0.01 and near inferior
conjunction (O star behind) on their orbital elements.
The new orbital elements of Marchenko et al., however, make a significant
difference to the expected configuration of the binary system, which is
shifted to about 45$^{\circ}$ after conjunction at the time of this
crucial IUE observation, so that it no longer usefully constrains the
inclination. Accordingly, we calculated $\psi$ for a range of inclinations
for the phase of each observation.

\begin{figure}
\vspace*{180pt}
\includegraphics{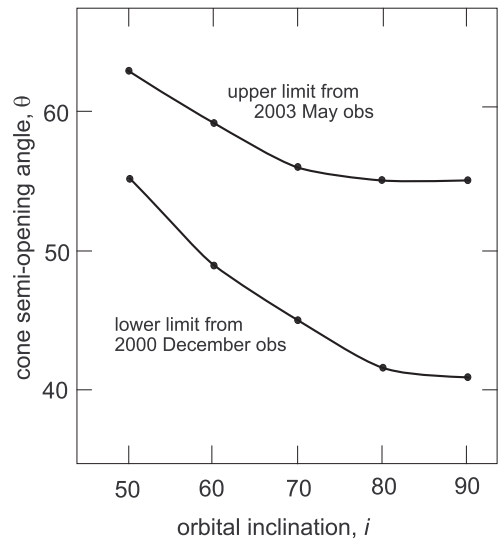}
\vspace*{45pt}
\caption{Limits on `cone' semi-opening angle $\theta$, as a function
of orbital inclination, $i$, set by the relative strengths of the
He\,{\sc i} P-Cygni absorption components in the 2000 December and
2003 spectra.}
\label{ThetaLim}
\end{figure}

The strong absorption in 2001 March and weak absorption in 2000 October
do not greatly constrain $\theta$, but the intermediate observations,
on 2000 December 25 and 26 and 2003 May, are particularly valuable.
The December spectra concur in showing an absorption feature as weak as
that in the October spectrum, indicating that we were still observing the
O star through the O star wind. This allows us to set a lower limit on
the `cone' angle $\theta$, which increases with falling inclination.
This is illustrated in Fig. {\ref{ThetaLim}}. Soon after the December
observations, the contact discontinuity between the winds swept through
our sightline, which then passed through the WC stellar wind, accounting
for the high absorption observed in the 2001 spectra when the `cone'
was directed away from us. By the time of the 2003 observation, the
He\,{\sc i} absorption was still strong, indicating that the `cone'
edge had not yet crossed our sightline, setting an upper limit on
$\theta$, again depending on the inclination (Fig. {\ref{ThetaLim}}).

Thus, $\theta$ is fairly tightly constrained, and our observations rule
out values around $36^{\circ}$ (Eichler \& Usov) derived from the
wind-momentum ratio, $\eta$, given by the mass-loss rates and terminal
velocities from W90.
This implies that the momentum of the O star wind had been underestimated
or that of the WC star overestimated. There is little scope for changing
the values $v_{\infty}$ for the two stars, which come from observations.
The mass-loss rate for the O star ($1.8 \times 10^{-6} M_{\odot}y^{-1}$)
was adopted from average values (Prinja et al.) while that of the WC star
$5.7 \times 10^{-5} M_{\odot}y^{-1}$ was derived from the 5-GHz flux density
at radio minimum, with allowance for the contribution from the wind of the
O star. We cannot rule out a $\sim$ three-fold higher mass-loss rate for
the O star but note that the value adopted by W90 is supported by recent
determinations ($1.63 \times 10^{-6} M_{\odot}y^{-1}$ for the O4~V((f*))
star HD 303308 and $1.28 \times 10^{-6} M_{\odot}y^{-1}$ for the O5~V((f))
star HD 15629, Repolust, Puls \& Herrero 2004). We therefore believe that
the WC star's mass-loss rate was overestimated because its wind is clumped.
A clumped wind with a volume filling factor $0.1$ (Dessart et al. 2000)
would have mass-loss rate a factor $\surd 10$ lower, yielding $\eta = 0.10$
and $\theta = 50^{\circ}$. This is consistent with the limits set by our
He\,{\sc i} observations, and will be adopted here.
                                                                                               
\subsection{The emission component of the He\,{\sc i} line}

The He\,{\sc i} line at 1.083 $\mu$m is a very good tracer of colliding-wind
phenomena in massive binaries (Stevens \& Howarth 1999). It forms in
the asymptotic region of the WR wind and thus it becomes sensitive to wind
interactions which can be observed from additional spectroscopic features
on its normally flat top. Eenens \& Williams (1994) observed a flat-topped
1.083 $\mu$m He\,{\sc i}-line profile from WR 140 with UKIRT and CGS2 at a
resolution of $R=637$ on 1990 June 20 (corresponding to orbital phase
$\phi $ = 0.66) although sub-peaks were observed from some other WC stars.
Re-observation of WR 140 with UKIRT on 1991 October 19 ($\phi $ = 0.83)
confirmed the absence of structure on the flat top of the 1.083-$\mu$m
line profile. Yet, by $\phi $ = 0.96, the phase of our first high-resolution
($R=4700$) UKIRT observation, a conspicuous sub-peak had appeared
at the short-wavelength end of the profile (Fig. \ref{wr1401083}).
We interpret the appearance of this new emission feature in the intervening
$0.13P$ as evidence of increased interaction of the WC and O type winds
as the stars moved closer to each other in their orbit.
Between phases 0.83 and 0.96, the separation of the stars decreased from
$1.28a$ to $0.5a$, so that the density of undisturbed winds having an
$r^{-2}$ density distribution would have increased by a factor of $6.5$.

The next high-resolution UKIRT spectra observed on 2000 December 25 and
26 show the sub-peak to have strengthened by a factor of $\simeq 2.5$
(Table \ref{velocity}). Corroboration of the presence of the sub-peak
comes from the $R=1000$ Mt. Abu $J$-band spectra (Fig. \ref{wr140jabu})
observed on December 21--23. The sub-peak was still present on the
short-wavelength end of the profile on 2001 January 2--3
(Fig. \ref{wr140jabu}), after which WR 140 was too close to the Sun for
us to observe. When WR 140 was
again observable in 2001 March, both the UKIRT high-resolution and the
Mt. Abu spectroscopy showed that the sub-peak had shifted to the red end
of the profile. We can also track the sub-peak in the $J$-band UKIRT
spectra (Fig. \ref{wr140jmk}): the sub-peak is visible at the red end of
the profile in the 2001 April--July spectra (apart from June 13, which has
poor S/N), weakening, broadening and moving back to the centre of the
profile in the August--November spectra. It may still be present in the
subsequent spectra, which do not show flat-topped profiles. By the time
of our next high-resolution spectrum, in 2003 May (Fig. \ref{wr1401083}),
the profile was flat-topped and the sub-peak had gone. The structure
on the 1.083-$\mu$m He\,{\sc i} profile is present for only a small
fraction of the period, but apparently a larger fraction than the
structures analagous to our sub-peaks on the 5696-\AA\ C\,{\sc iii}
and 5896\AA\ He\,{\sc i} lines observed by Marchenko et al.
                                                                                               
Marchenko et al. found that the strengths of the structures on the
5696-\AA\ C\,{\sc iii} and 5896\AA\ He\,{\sc i} lines varied with
separation of the binary components more steeply than the $1/d$
dependency expected. We have fewer measurements, but it seems that
the 1.083-$\mu$m behave similarly. As noted above, the flux in the
sub-peak increased by a factor $\simeq 2.4$ between our 2000 October
and December observations while the reciprocal separation, $a/d$,
increased by a factor of 2.05. These spectra were observed before
dust formation began and were calibrated with lower resolution
spectrum and photometry. The flux calibration of the 2001 March
spectra, observed when there was a contribution from heated dust to
the continuum, is less certain,  being based on a magnitude, $J=5.26$,
taken for these phases from the J band light curve of W90 and corrected
for the thermal emission from the dust. But it seems that the flux
in the sub-peak is stronger still in 2001 March.


The movement of the sub-peak is readily understood, qualitatively at
least, in terms of the changing orientation of the wind-collision zone
with orbital motion, especially around the time of periastron passage.
Fig. {\ref{wr140orbit}} shows the configuration of WR 140 with the WC
star at the origin and the O star in its relative orbit at the orbital
phases of the high-resolution UKIRT spectra.  The shock cone, wrapped
around the O star, is also shown. The observer is in the negative Y
direction (i.e. the line of nodes is parallel to the X axis).
The high resolution UKIRT spectra of the 1.083 $\mu$m He\,{\sc i} line
are overplotted on this diagram to demonstate the evolution of the sub-peak
on the He\,{\sc i} line. Its position shifted in accordance with the
motion of the components and the associated orientation of the shock cone
with respect to the observer, and intensity, with the separation between
the components.

We can develop a simple model assuming that the emitting material flows
along a thin shell defined by the contact discontinuity, but recognise
that real wind collisions are turbulent, full of instabilities and much
more complicated (e.g. Walder \& Folini 2002, Folini \& Walder 2002).
In Fig.~{\ref{wr140oct}}, showing a view of the configuration of the system
in 2000 October, the shaded area shows the compressed wind material flowing
in a thin shell along the contact discontinuity. Cant\'o et al. (1996) have
given a formula for the wind velocity $v_s$ as a function of position on the
surface as it accelerates from near the stagnation point between the stars
to reach an asymptotic value on the asymptotic region of the contact
discontinuity, defined by $\theta_{1\infty}$ subtended at the WC star in
the notation of Cant\'o et al., approximated by the `cone' of opening
semi-angle $\theta$ in the discussion in the previous section. We can
resolve $v_s$ into components parallel ($v_{axis}$) and perpendicular
($v_z$) to the axis of symmetry. Emission by material flowing in this
shell at any particular phase would have radial velocities in the range:
\begin{equation}
RV = -v_{axis} cos(\psi) \pm v_z sin(\psi)
\label{RVdef}
\end{equation}
\noindent where $\psi$ for that phase come from Eqn \ref{psidef}. We
do not know where along the surface the emission arises from.  If the
material near the stagnation point is shock-heated to $\sim 10^7$ K, it
seems reasonable to assume that the He recombination occurs down-wind
and that the emission arises from material on the asymptotic or `cone'
region of the surface, where the velocity has reached its asymptotic,
constant value.
This becomes analogous to the thin shell model of L\"uhrs (1997)
and Hill, Moffat \& St-Louis (2002) except that, instead of solving
for the cone angle and flow velocity, we will adopt the value of
$\theta$ derived above and determine the flow velocity following
Cant\'o et al. (eqn 29) from the terminal velocities of the WC and
O stars (2860 km s$^{-1}$ and 3100 km s$^{-1}$ respectively, W90)

\begin{figure}
\vspace*{180pt}
\includegraphics{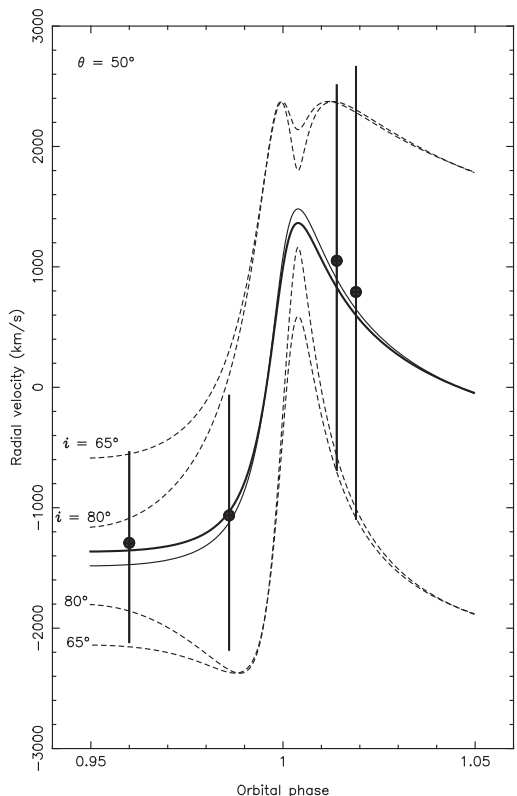}
\vspace*{135pt}
\caption{Comparison of variation with orbital phase of observed radial
velocities of sub-peaks on the 1.083-$\mu$m line (vertical bars at phases
of observation show extents of the sub-peaks, with $\bullet$ marking the
central velocity) with the model described in the text for two values
of orbital inclination. The solid line shows the central velocity, the
thicker line for $i=65^{\circ}$, and the thinner line for $i=80^{\circ}$.
The extents of the emission feature for the two values of $i$ are marked
with dashed lines, labelled where the model is sensitive to inclination.}
\label{model}
\end{figure}

From these, we derived velocity components $v_{axis} = 1505$ km$^{-1}$
and $v_{z} = 1835$ km$^{-1}$ for our $\eta = 0.10$ surface, which
corresponds to $\theta = 50^{\circ}$. We use these and Eqn \ref{RVdef}
to model the variation of the RV of the centre and extent of the
emission from the thin shell as a function of phase for different
values of the inclination. This is shown for $i = 80^{\circ}$ and
$i = 65^{\circ}$ in Fig. \ref{model}, where the models are compared
with the observed velocities of the centre and extent measured from
our high-resolution UKIRT spectra.

The agreement is very encouraging: the simple models successfully
explains the movement of the centre of the line and reasonably approximate
its width. From comparision with the October observation, we suggest that
the orbital inclination is close to $i=65^{\circ}$.
The discrepanicies at the blue edges of the pre-eclipse sub-peaks are
expected since we are not able to measure accurately the `real' blue edge
of the sub-peak before the periastron passage since it overlaps the steep,
P-Cygni absorbed, region of the profile, which also results in an 
under-estimation of the flux in the subpeak during these phases. Given 
the uncertainties, we have not attempted a formal solution for inclination. 
Tests showed that the model is not very sensitive to $\theta$ in the range 
set by the absorption components (Fig. \ref{ThetaLim}).

\section{Discussion}

The flux in the sub-peak (Table \ref{velocity}) on the He\,{\sc i} 
profile is a significant source of cooling for the shock-heated material, 
exceeding the 2--6 keV X-ray flux  ($2.5 \times 10^{-14}$ Wm$^{-2}$) 
observed with EXOSAT near the time of the 1985 periastron passage (W90) 
or the 1--10 keV flux ($2.4 \times 10^{-14}$ Wm$^{-2}$) observed with 
ASCA by Koyama et al. (1994) in 1993 June, shortly after the following 
periastron passage.  

Koyama et al. (1994) also noted that their X-ray flux was not significantly 
different from that observed with Ginga in 1987 (Koyama et al. 1990), at a 
phase when the orbital separation was $\simeq 4$ times that of their 1993 
observation. They pointed out that this, and the EXOSAT results, showed no 
convincing evidence that the X-ray flux above 2 keV, which should be 
relatively unaffected by interstellar or orbitally modulated circumstellar 
extinction, varied systematically with orbital separation. This was in 
contrast to the $1/d$-dependence expected theoretically for the X-ray 
luminosity of WR140 (Stevens, Blondin \& Pollock 1992), on the basis that 
the shocked winds were close to adiabatic because the cooling time of the 
shocked X-ray emitting material was long compared with the characteristic 
flow time.

This discrepancy was studied further with additional ASCA data by Zhekov \& 
Skinner (2000) and Pollock, Corcoran \& Stevens (2002). Both studies sought 
to adjust the orbital elements of WR 140 derived by W90 to recover the 
$1/d$-dependence for the X-ray fluxes, either reducing the eccentricity to 
$e=0.55$ or bringing forward T$_0$ by 0.025P (72 days). The proposed lower 
eccentricity is effectively ruled out by confirmation of the high 
eccentricity in the more recent orbital solutions by Setia Gunawan et al. 
($e=0.87$) and Marchenko et al. ($e=0.881$). 
From the variations of the 1.083$\mu$m He\,{\sc i} line profile observed 
around the periastron passage (Figs {\ref{wr140jmk}}, {\ref{wr140jabu}} and 
{\ref{wr1401083}}), we see that the sub-peak remained blue shifted until 2001 
January 2 and had moved to the red end of the profile by March 18. If we are 
correct in interpreting this movement as a change of the orientation of the 
wind-collision region moving with the stars in their orbits, then quadrature 
and periastron passage must have occurred after January 2. This is consistent 
with the predictions of T$_0$ of 2001 February 20 (W90) or February 6 
(from elements by Marchenko et al.) but rules out the 72-day shift to 
December 10 proposed by Zhekov \& Skinner. (Setia Gunawan et al. also 
proposed moving T$_0$ forward, by 104 days. relative to the elements of W90, 
but with a relatively large uncertainty, 117 days, which does not constrain 
our solution.)

We propose instead that the X-ray flux departs from a $1/d$-dependence 
because the shocked WC wind is not totally adiabatic during periastron passage 
as a consequence of significant additional cooling in the He\,{\sc i} 
sub-peak. 

\section{Conclusions}

Infrared spectroscopy of WR 140 during and after its 2001 periastron 
passage showed dilution of the emission lines consistent with the 
formation of circumstellar dust between 2001 January 3 and March 26, 
phases 0.989 and 0.017. The He\,{\sc i} 1.083-$\mu$m line showed a 
strong P-Cygni profile. Variation of the strength of the absorption 
component measured from high-resolution spectra was interpreted in 
terms of the passage of different parts of the colliding wind 
structure in our line of sight to the two stars. This allowed us to 
set fairly tight limits on the opening half angle of the `cone' 
modelling the interaction region: $\theta = 50^{\circ}\pm 8^{\circ}$. 
This, in turn, indicated a wind-momentum ratio $\eta = 0.10$, 
signifcantly higher than the $\eta = 0.034$ found from the mass-loss 
rates and terminal velocities of the WC7 and O components. 
We suggested  that the mass-loss rate of the WC7 star, 
$5.7 \times 10^{-5} M_{\odot}y^{-1}$ derived from the 5-GHz flux 
density at radio minimum was overestimated because the wind is clumped. 

At phases ($\Delta\phi \simeq 0.1P$) close to periastron passage, the 
normally flat-topped emission component of the 1.083-$\mu$m line 
showed the appearance of a sub-peak which moved from the blue to the 
red end of the profile during periastron. The evolution of the position 
and width of the sub-peak were found to be reasonably conistent with 
the flow of the emitting material along the surface of the interaction  
region modelled by a cone having $\theta = 50^{\circ}$. The maximum 
radiative flux in the sub-peaks is greater than the 1--10 keV X-ray 
flux at periastron and is a significant source of cooling of the shocked
WC wind. We suggest that the shocked WC wind is not totally adiabatic 
near periastron, thereby accounting for the departure of the X-ray 
flux from the previously expected $1/d$-dependence and, as suggested 
by Marchenko et al., allowing dust formation.

\section*{Acknowledgments}
We would like to thank UKIRT Service Observing program for obtaining some 
of the $JHK$ spectra of WR 140. UKIRT is operated by the Joint Astronomy
Centre, Hilo, Hawaii,  on behalf of the U.K. Particle Physics and Astronomy
Research Council.  Mt. Abu observatory is operated by the Physical Research
Laboratory, Ahmedabad, India, funded by the Dept. of Space, Govt. of India.
We also would like to thank the staff members of UKIRT and Mt. Abu
observatories, who helped us with data collection at various stages of this 
monitoring campaign. WPV would like to thank Andrew J. Adamson for his 
help and support for this work.  We also thank Sean Dougherty and 
referee Paul Crowther for many valuable comments and suggestions.

\label{lastpage}

\end{document}